\documentclass[10pt]{article}
\pdfoutput=1
\usepackage{amssymb,amsmath,mathrsfs,enumerate}
\usepackage{graphicx,rotate,multicol,wrapfig}
\usepackage[margin=10pt,labelfont=bf]{caption}
\usepackage{cite}
\usepackage[colorlinks=true,
            linkcolor=red,
            urlcolor=blue,
            citecolor=blue]{hyperref}

\usepackage{color}
\usepackage{tikz,braket}
\usetikzlibrary{arrows,positioning, shapes.geometric}
\long\def\rpl#1!!#2!!{\textcolor{red}{#1} \textcolor{blue}{#2}}

\long\def\rpl#1!!#2!!{\textcolor{red}{#1} \textcolor{blue}{#2}}

\def \order(#1){{\cal O} \left(#1 \right)}
\DeclareMathOperator{\Tr}{Tr}

\evensidemargin=\oddsidemargin
\def\Eqn#1{Eq.\ (\ref{#1})}
\def\Eqs#1#2{Eqs.\ (\ref{#1}) and (\ref{#2})}
\allowdisplaybreaks

\begin{document}

\hfill {\small  FTUV-16-0426, IFIC-16-14}\\[0.5cm]

\begin{center}
	{\Large \bf Updated scalar sector constraints in Higgs triplet model} \\
	\vspace*{1cm} {\sf Dipankar	Das\footnote{dipankar.das@uv.es}, ~Arcadi Santamaria\footnote{arcadi.santamaria@uv.es}} \\
	\vspace{10pt} {\small\em Departament de F\'{i}sica T\`{e}orica, Universitat de Val\`{e}ncia and IFIC, Universitat de Val\`{e}ncia-CSIC  \\
	Dr. Moliner 50, E-46100 Burjassot (Val\`{e}ncia), Spain}
	
	\normalsize
\end{center}

\begin{abstract}
We show that in the Higgs triplet model, after the Higgs discovery, the mixing angle in the CP-even sector can be strongly constrained from unitarity. We also discuss how
large quantum effects in $h\to\gamma\gamma$ may arise in a SM-like scenario and a certain part of the parameter space can be ruled out
from the diphoton signal strength. Using $T$-parameter and diphoton signal strength measurements, we  update the bounds on the
nonstandard scalar masses.
\end{abstract}

\bigskip
\section{Introduction}
\label{s:intoduction}
Neutrino masses are one of the main motivations we have at present for physics beyond the Standard Model~(SM). Although a minimal extension of the SM by adding three right-handed neutrinos with Dirac mass terms is still allowed by all neutrino data, this is not the preferred scenario for it does not explain the smallness of neutrino masses or why lepton number should exactly be conserved. 

Scenarios in which the smallness of neutrino masses is linked to the non-conservation of lepton number are usually considered more natural. The simplest versions of these scenarios are realized at tree level and are known under the name of the seesaw mechanisms. Seesaws of type I \cite{Minkowski:1977sc,Ramond:1979py,GellMann:1980vs,Yanagida:1979as,Mohapatra:1979ia} and III \cite{Foot:1988aq,Ma:2002pf} extend the SM with new fermions, singlet and triplet respectively, with a Majorana mass term, $M$, which breaks explicitly lepton number.  On the other hand, seesaw of type II \cite{Konetschny:1977bn,Cheng:1980qt,Lazarides:1980nt,Magg:1980ut,Schechter:1980gr} adds only a scalar triplet with hypercharge~2. In this case, lepton number is broken explicitly in the scalar potential by a trilinear coupling, $\mu$ (see also \cite{Gelmini:1980re,Georgi:1981pg} for a variation with lepton number broken spontaneously).

Seesaws of type I and III give neutrino masses of order $m_\nu \sim y v_d^2/M$ with $y$ being the Yukawa coupling and $v_d$ the vacuum expectation value (VEV) of the SM doublet. Then, if $M\gg v_d$, masses can naturally be small. However, the same parameters that appear in the neutrino masses also appear in the mixing of new fermions with ordinary fermions, which is proportional to $y v_d/M$ and must be very small. This makes these types of models very difficult to test. 

Seesaw of type II, however, generates neutrino masses of order $m_\nu \sim y \mu v_d^2/M^2$, where now $y$ is the Yukawa coupling of the scalar triplet to lepton doublets and $M$ the triplet mass. The trilinear coupling $\mu$ is protected by symmetry and can be naturally small. This allows for small neutrino masses compatible with a rich phenomenology (the new scalars of the model could be produced at the LHC \cite{Cuypers:1996ia,delAguila:2008cj,Perez:2008ha,Melfo:2011nx,Chun:2012zu,delAguila:2013mia,Chun:2013vma,delAguila:2013yaa} and there could be lepton flavor violating processes like $\mu \rightarrow 3e$ or $\mu\rightarrow e\gamma$ \cite{Leontaris:1985qc,Pich:1985uv,Bernabeu:1987dz,Bilenky:1987ty,Akeroyd:2009nu}), raising the possibility to test the mechanism of neutrino masses in non-oscillation experiments. 

In an effort to fixate the scalar spectrum of the model, in this article we revisit the scalar potential of the type II seesaw model taking into account unitarity and the stability of the potential. Moreover, on the phenomenological side, we will also consider the constraints coming from the oblique $T$-parameter and Higgs decay branching ratios, in particular $h\rightarrow \gamma\gamma$, which are largely independent on the other sectors of the theory (Yukawa couplings and neutrino masses). In fact, the one-loop $T$-parameter only depends, to a good approximation,  on the mass splitting between the scalars, while the new contributions to $h\rightarrow \gamma\gamma$ can be expressed  in terms of the scalar masses and the mixing angle in the neutral scalar sector. 

A very important ingredient of our analysis is the requirement that the triplet scalar VEV, $v_t$, is much smaller than the electroweak scale. This is, in part, required by the tree-level $\rho$-parameter and in part, by the requirement of small neutrino masses. It has been noticed, in this case, that the some relations involving the scalar masses can be obtained and that the $T$-parameter constrains the splitting  to be less that about $50$~GeV (see for instance \cite{Melfo:2011nx}). We will re-derive these constraints by using perturbative unitarity and the $T$-parameter. After this, the spectrum of scalar masses is practically fixed up to a global scale and a small splitting between masses and can be expressed in terms of three parameters (apart from the known parameters, the Higgs mass and the electroweak scale, and the triplet VEV which becomes irrelevant for $v_t< 1$~GeV). Then, we will derive a lower bound on the global scale by adding present data on $h\rightarrow \gamma\gamma$.

Similar analyses have been performed several times in the literature \cite{Arhrib:2011uy,Chun:2012jw,Arbabifar:2012bd,Dev:2013ff}. But we differ from those in the sense that we express our results in terms of the experimentally measurable quantities. In the process, we have been able to unravel some features which, we believe, have not been emphasized before. Additionally we also point out that the current measurement of the diphoton signal strength enables us to put lower limits on the nonstandard scalar masses, which are competitive or, in some cases, better than the direct search limits. 

In Section \ref{s:potential} we study the scalar potential, the scalar mass spectrum and the constraints from unitarity and stability. In Section \ref{s:analysis} we perform a complete numerical analysis by including also the constraints from the $\rho$-parameter. In Section \ref{s:hgg} we study the impact of the data on $h\rightarrow \gamma\gamma$ and obtain the lower bound on the masses. Finally in Section \ref{s:conclusions} we present our conclusions.

\section{The scalar potential}
\label{s:potential}
The Type~II seesaw model extends the Higgs sector of the standard model by adding one scalar $SU(2)_L$ triplet~($\Delta$) with hypercharge,
$Y_\Delta=2$. The most general scalar potential involving this triplet and the standard $SU(2)_L$ doublet, $\Phi$, is given by\cite{Arhrib:2011uy}
\begin{eqnarray}
\label{e:potential}
V &=& -m_{\Phi}^2\left(\Phi^\dagger\Phi\right) +M^2\Tr\left(\Delta^\dagger\Delta\right) +\left\{\mu \left(\Phi^Ti\sigma_2\Delta^\dagger\Phi\right) + {\rm h.c.} \right\} +\frac{\lambda}{4}\left(\Phi^\dagger\Phi\right)^2
+\lambda_1\left(\Phi^\dagger\Phi\right)\Tr\left(\Delta^\dagger\Delta\right) \nonumber \\
&& +\lambda_2\left\{\Tr\left(\Delta^\dagger\Delta\right)\right\}^2 +\lambda_3\Tr\left[\left(\Delta^\dagger\Delta\right)^2\right] +\lambda_4\left(\Phi^\dagger\Delta \Delta^\dagger\Phi \right) \,,
\end{eqnarray}
where `$\Tr$' represents the trace over $2\times2$ matrices and $\sigma_2$ is the second Pauli matrix. We can take all the
parameters in the potential to be real without any loss of generality\cite{Gunion:1989ci,Dey:2008jm}. Denoting by $v_d$ ($v_t$)  the VEV of the doublet (triplet) scalar field, the minimization conditions read
\begin{subequations}
\label{e:min}
\begin{eqnarray}
m_{\Phi}^2&=& \frac{\lambda v_d^2}{4} +\frac{(\lambda_1+\lambda_4)v_t^2}{2} -\sqrt{2}\mu v_t \,,  \\
M^2 &=& -(\lambda_2+\lambda_3)v_t^2 - \frac{(\lambda_1+\lambda_4)v_d^2}{2} +\frac{\mu v_d^2}{\sqrt{2}v_t} \,.
\label{M2}
\end{eqnarray}
\end{subequations}
After the spontaneous symmetry breaking, we represent the scalar multiplets in the following way:
\begin{eqnarray}
\Phi = \frac{1}{\sqrt{2}} \begin{pmatrix} \sqrt{2} w_d^+ \\ v_d+h_d+iz_d \end{pmatrix} \,, &&
\Delta = \frac{1}{\sqrt{2}} \begin{pmatrix}  w_t^+ & \sqrt{2}\delta^{++} \\ v_t+h_t+iz_t & -w_t^+ \end{pmatrix} \,,
\end{eqnarray}
and the electroweak VEV is then given by,
\begin{eqnarray}
v = \sqrt{v_d^2+2v_t^2} = 246~{\rm GeV} \,.
\end{eqnarray}
From the observed value of the electroweak $\rho$-parameter, $v_t$ is expected to be $\order(1~{\rm GeV})$ or less\cite{Georgi:1981pg,Perez:2008ha,Melfo:2011nx,Arhrib:2011uy,Kanemura:2012rs}. 

\subsection{Physical eigenstates}
Using \Eqn{e:min} we first trade $m_{\Phi}^2$ and $M^2$ for $v_d$ and $v_t$ in the potential of \Eqn{e:potential}.
The fields, $\delta^{\pm\pm}$, represent the doubly charged  scalar with mass
\begin{eqnarray}
m_{++}^2 = \frac{\mu v_d^2}{\sqrt{2}v_t} -\frac{\lambda_4}{2}v_d^2 -\lambda_3v_t^2 \,.
\end{eqnarray}
The mass squared matrix in the singly charged sector can be rotated to the physical basis through
\begin{eqnarray}
\begin{pmatrix} \omega^\pm \\ H^\pm \end{pmatrix} = \begin{pmatrix} \cos\beta & \sin\beta \\ -\sin\beta & \cos\beta \end{pmatrix} \begin{pmatrix} w_d^\pm \\ w_t^\pm \end{pmatrix} \,, ~~~ {\rm with} ~~ \tan\beta = \frac{\sqrt{2}v_t}{v_d} \,,
\end{eqnarray}
to obtain the charged Goldstone~($\omega^\pm$) along with a  singly charged scalar~($H^\pm$) with
mass
\begin{eqnarray}
m_{+}^2= \frac{(2\sqrt{2}\mu-\lambda_4v_t)}{4v_t}(v_d^2+2v_t^2) \,.
\end{eqnarray}
The assumption of real VEVs allows us to define electrically neutral mass eigenstates which are also eigenstates of $CP$.
The mass squared matrix in the $CP$-odd sector can be rotated to the physical basis through
\begin{eqnarray}
\begin{pmatrix} \zeta \\ A \end{pmatrix} = \begin{pmatrix} \cos\beta' & \sin\beta' \\ -\sin\beta' & \cos\beta' \end{pmatrix} \begin{pmatrix} z_d \\ z_t \end{pmatrix} \,, ~~~ {\rm with} ~~ \tan\beta' = \frac{2v_t}{v_d} \,,
\end{eqnarray}
to obtain the neutral Goldstone~($\zeta$) along with a pseudoscalar~($A$) with mass
\begin{eqnarray}
m_{A}^2= \frac{\mu}{\sqrt{2}v_t}(v_d^2+4v_t^2) \,.
\end{eqnarray}
Finally, for the $CP$-even part we have:
\begin{subequations}
\begin{eqnarray}
M_S^2 &=& \begin{pmatrix}
 A_S & -B_S \\  -B_S &  C_S
\end{pmatrix} \,, \\
{\rm where,} ~~~ A_S &=& \frac{\lambda v_d^2}{2} \,, \\
B_S &=& \sqrt{2}\mu v_d -(\lambda_1+\lambda_4)v_tv_d \,, \\
C_S &=& \frac{\mu v_d^2}{\sqrt{2}v_t}+2(\lambda_2+\lambda_3)v_t^2 \,.
\end{eqnarray}
\end{subequations}
We can obtain the physical eigenstates through the following rotation:
\begin{subequations}
\begin{eqnarray}
\label{e:alpha}
\begin{pmatrix} h \\ H \end{pmatrix} = \begin{pmatrix} \cos\alpha & \sin\alpha \\ -\sin\alpha & \cos\alpha \end{pmatrix} \begin{pmatrix} h_d \\ h_t \end{pmatrix} \,, && ~ {\rm with,} ~~ \tan2\alpha = \frac{2B_S}{C_S-A_S} =\frac{\sqrt{2}\mu v_d -(\lambda_1+\lambda_4)v_tv_d}{\frac{\mu v_d^2}{2\sqrt{2}v_t}+(\lambda_2+\lambda_3)v_t^2 -\frac{\lambda v_d^2}{4}} \,, \\
{\rm with~ masses,} && m_h^2 = (A_S+C_S)-\sqrt{(A_S-C_S)^2+4B_S^{2}} \,, \\
  && m_H^2 = (A_S+C_S)+\sqrt{(A_S-C_S)^2+4B_S^{2}} \,.
\end{eqnarray}
\end{subequations}

\paragraph{SM like limit:} Even with $v_t \ll v_d$, to make the tree level couplings of the lightest $CP$-even scalar~($h$)
close to those in the SM, we need to set $\sin\alpha=0$. 
From \Eqn{e:alpha} we see that there are two different ways to obtain this limit as discussed below:
\subparagraph{Case~I:} The first option is to invoke a fine tuning so that the numerator
in the expression for $\tan2\alpha$ vanishes. The condition reads:
\begin{subequations}
\label{e:case1}
\begin{eqnarray}
&& \sqrt{2}\mu = (\lambda_1+\lambda_4)v_t \,, \\
&\Rightarrow& \sin\alpha = 0 \,, ~~{\rm with} ~ v_t \ne 0 \,. \label{case1}
\end{eqnarray}
\end{subequations}
Note that, in this case, we can make the tree level couplings of $h$ to be close to SM {\em without} demanding
$v_t$ to be exactly zero. This limit is interesting in view of the fact that we can still have small neutrino masses through a
small value for $v_t$ and at the same time be very close to the SM. But later we will show that the charged scalars, in this limit, can contribute substantially in the diphoton decay amplitude
and therefore this limit is {\em ruled out}.
\subparagraph{Case~II:} In a second and more conventional approach, the SM-like
limit is obtained as a direct consequence of $v_t$ being arbitrarily small. To be more precise, in the limit 
and $M^2 \gg v^2$, we can approximate \Eqn{M2} as
\begin{eqnarray}
\label{e:M2}
v_t \approx \frac{\mu v^2}{\sqrt{2}M^2} \,.
\end{eqnarray}
In this case the expression for $\tan2\alpha$ in \Eqn{e:alpha} can be simplified into
\begin{eqnarray}
 \tan2\alpha \approx \frac{4v_t}{v_d} \,, ~~ \Rightarrow \sin\alpha \approx \frac{2v_t}{v_d} \,, ~~{\rm with} ~ v_t \rightarrow 0 \,. \label{case2}
\end{eqnarray}
As we will show later, this limit corresponds to the decoupling of heavy nonstandard scalars.

It is now instructive to count the number of free parameters in the scalar potential.  Note that, \Eqn{e:potential} contains eight free parameters. As mentioned before, $m_{\Phi}^2$ and $M^2$ can be traded in favor of $v_d$ and $v_t$. The
remaining six parameters can be exchanged for five physical masses and the mixing angle, $\alpha$, as follows\cite{Arhrib:2011uy}:
\begin{subequations}
\label{e:inverted}
\begin{eqnarray}
\mu &=& \frac{\sqrt{2}m_A^2v_t}{v_d^2+4v_t^2} \,, \label{e:mu} \\
\lambda &=& \frac{2}{v_d^2}\left(m_H^2\sin^2\alpha+m_h^2\cos^2\alpha\right) \,, \\
\lambda_1 &=& \frac{4m_+^2}{v_d^2+2v_t^2} - \frac{2m_A^2}{v_d^2+4v_t^2} -\frac{\sin\alpha\cos\alpha}{v_dv_t}
\left(m_H^2-m_h^2 \right) \,, \\
\lambda_2 &=& \frac{1}{v_t^2}\left[ \frac{1}{2}\left(m_h^2\sin^2\alpha+m_H^2\cos^2\alpha\right) + \frac{v_d^2m_A^2}{2(v_d^2+4v_t^2)}- \frac{2v_d^2m_+^2}{v_d^2+2v_t^2} + m_{++}^2\right] \,, \\
\lambda_3 &=& \frac{1}{v_t^2}\left[ \frac{2v_d^2m_+^2}{v_d^2+2v_t^2} - \frac{v_d^2m_A^2}{v_d^2+4v_t^2}- m_{++}^2\right] \,, \\
\lambda_4 &=& \frac{4m_A^2}{v_d^2+4v_t^2} - \frac{4m_+^2}{v_d^2+2v_t^2} \,.
\end{eqnarray}
\end{subequations}
Among the eight redefined parameters that appear on the RHS of \Eqn{e:inverted}, not all are unknown.  We already
know $v= \sqrt{v_d^2+2v_t^2}= 246$~GeV and under the assumption that the lightest CP-even Higgs is what has been 
found at the LHC, $m_h\approx 125$~GeV is also known. The compatibility of Higgs signal strengths into different
decay channels with their corresponding SM expectations tells us to focus near the SM-like limit, $\sin\alpha\approx0$.
We will see how the smallness of $v_t$ in association with unitarity and stability entail strong correlations among the
remaining four nonstandard masses, $\{m_H,m_A,m_+,m_{++}\}$, making the scalar potential of Type~II seesaw model
constrained very strongly.

\subsection{Theoretical constraints from vacuum stability and unitarity}
We need to ensure that there is no direction in the field space along which the potential becomes
infinitely negative.  The  conditions for the potential of \Eqn{e:potential} to be bounded from below read\cite{Bonilla:2015eha}
\begin{subequations}
\label{e:stability}
\begin{eqnarray}
&& \lambda \ge 0 \,, ~~\lambda_2+\lambda_3 \ge 0 \,, ~~ \lambda_2+\frac{\lambda_3}{2} \ge 0 \,, ~~ \lambda_1+\sqrt{\lambda(\lambda_2+\lambda_3)}  \ge 0 \,, ~~ \lambda_1+\lambda_4+\sqrt{\lambda(\lambda_2+\lambda_3)} \ge 0 \,, \\
{\rm and,} &&  \left[\left|\lambda_4 \right|\sqrt{\lambda_2+\lambda_3}-\lambda_3\sqrt{\lambda}\ge 0 \,, ~~{\rm or,}~~
2\lambda_1+\lambda_4+\sqrt{\left(2\lambda\lambda_3-\lambda_4^2 \right)\left(\frac{2\lambda_2}{\lambda_3}+1 \right) \ge 0} \right] \,.
\end{eqnarray}
\end{subequations}
The $S$-matrix eigenvalues that will be constrained from unitarity of the scattering amplitudes are also listed
below\cite{Arhrib:2011uy}:
\begin{subequations}
\label{e:unieigen}
\begin{eqnarray}
&& \left|(\lambda+4\lambda_2+8\lambda_3)\pm \sqrt{(\lambda-4\lambda_2-8\lambda_3)^2+16\lambda_4^2} \right| \le 64\pi \,, \\
&&\left|(3\lambda+16\lambda_2+12\lambda_3)\pm \sqrt{(3\lambda-16\lambda_2-12\lambda_3)^2+24(2\lambda_1+\lambda_4)^2} \right| \le 64\pi \,, \\
&& \left|\lambda \right| \le 32\pi \,, \label{e:u3} \\
&& \left|2\lambda_1+3\lambda_4 \right| \le 32\pi \,, \\
&& \left|2\lambda_1-\lambda_4 \right| \le 32\pi \,, \\
&& \left|\lambda_1 \right| \le 16\pi \,, \\
&& \left|\lambda_1+\lambda_4 \right| \le 16\pi \,, \label{e:u7} \\
&& \left|2\lambda_2-\lambda_3 \right| \le 16\pi \,, \label{e:u8} \\
&& \left|\lambda_2 \right| \le 8\pi \,, \label{e:u9} \\
&& \left|\lambda_2+\lambda_3 \right| \le 8\pi \,. \label{e:u10}
\end{eqnarray}
\end{subequations}

\section{Numerical analysis and results}
\label{s:analysis}
As a first level of simplification, we can use \Eqn{e:inverted} into the inequality (\ref{e:u10}) and remembering $\lambda_2+\lambda_3\ge 0$, we may write
\begin{eqnarray}
0\le \left(m_h^2\sin^2\alpha +m_H^2\cos^2\alpha \right) -\frac{v_d^2m_A^2}{v_d^2+4v_t^2} \le 16\pi v_t^2 \,.
\end{eqnarray}
Therefore, for $v_t < \order(1~{\rm GeV})$, we can very well approximate
\begin{eqnarray}
m_A^2 &\approx& m_h^2\sin^2\alpha +m_H^2\cos^2\alpha  \label{e:mA1} \,.
\end{eqnarray}
%
This automatically implies $m_h< m_A < m_H$.
Similar considerations for the inequalities (\ref{e:u8}) or (\ref{e:u9}) enable us to express the doubly charged scalar mass
as follows:
\begin{eqnarray}
m_{++}^2 \approx 2m_+^2-m_A^2 &\approx& 2 m_{+}^2 -\left(m_h^2\sin^2\alpha +m_H^2\cos^2\alpha \right) \label{e:mCC1}  \,.
\end{eqnarray}
%
We will exemplify our numerical results by setting $v_t=1$~GeV. We also take $v=246$~GeV and $m_h=125$~GeV as input parameters. Then we perform random scan over the $\{\sin\alpha, m_H, m_{+} \}$ space by varying the parameters within the following ranges:
\begin{eqnarray}
\sin\alpha \in [-0.2,0.2] \,, ~~ m_H \in [125, 2000]~{\rm GeV}\,, ~~ m_{+}\in [0,2000]~{\rm GeV}\,,
\end{eqnarray}
and calculate $m_A$ and $m_{++}$ through the relations (\ref{e:mA1}) and (\ref{e:mCC1})\footnote{Nonzero values of $\lambda_2$ and $\lambda_3$ will cause negligible deviations from \Eqs{e:mA1}{e:mCC1}. We have taken this effect into account in our numerical analysis.}.  In anticipation that the Higgs data will continue to agree with the SM with increasing accuracy  in  the  upcoming runs of the LHC, we vary $\sin\alpha$ in a rather narrow
range around $\sin\alpha=0$.  Next we compute the $\lambda_i$s
using \Eqn{e:inverted} and check whether the unitarity and stability conditions, given in \Eqs{e:unieigen}{e:stability}, are
satisfied. Note that, since we are scanning in terms of the physical parameters, positivity of the masses are guaranteed and therefore, we are in a local minimum. Moreover, we have explicitly checked that $V_{\rm min}<0$ for every point in our scan. We discuss below the results of our analysis.

\begin{figure}[htbp!]
\begin{minipage}{0.46\textwidth}
\centerline{\includegraphics[width=8cm,height=6cm]{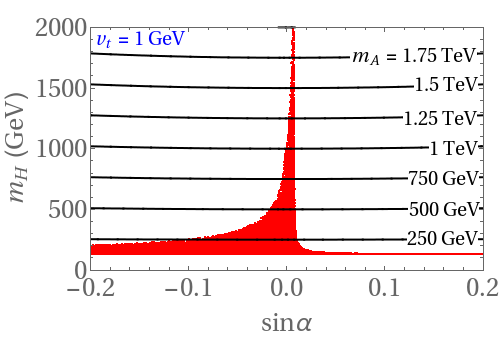}}
\caption{\em Allowed points in $\sin\alpha$-$m_H$ plane from unitarity and stability for $v_t=1$~GeV. The continuous lines
are contours for $m_A$ drawn using \Eqn{e:mA1}.}
\label{f:samH}
\end{minipage}
\hfill
\begin{minipage}{0.46\textwidth}
\centerline{\includegraphics[width=8cm,height=6cm]{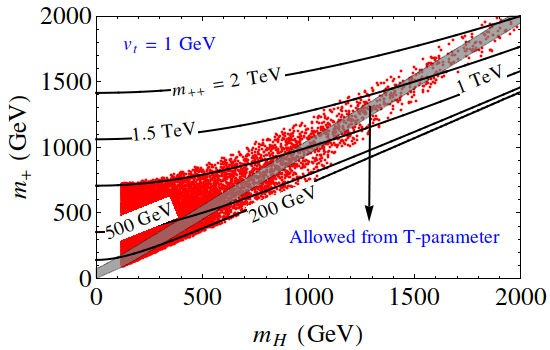}}
\caption{\em Allowed points in $m_H$-$m_+$ plane from unitarity and stability for $v_t=1$~GeV. The continuous lines
are contours for $m_{++}$ drawn using \Eqn{e:mCC1} for $\sin\alpha\approx0$. The allowed region from $T$-parameter has also been
shaded assuming $m_H\approx m_A$.}
\label{f:mHmC}
\end{minipage}
\end{figure} 

From Fig.~\ref{f:samH} we see that a heavy $m_H$ requires $\sin\alpha\approx 0$, {\it i.e.}, we are automatically pushed towards
a SM-like situation for small $v_t$ and heavy nonstandard scalars. However a closer inspection of Fig.~\ref{f:samH} reveals
that the peak does not occur precisely at $\sin\alpha= 0$ but at a small non negative value of $\sin\alpha$. The reason for this can
be understood from the unitarity condition~(\ref{e:u7}) which, in terms of the physical masses reads
\begin{eqnarray}
\label{e:peak}
\left|m_h^2\sin^2\alpha+m_H^2\cos^2\alpha -\sin\alpha\cos\alpha \frac{v_d}{2v_t}\left(m_H^2-m_h^2\right) \right| \le 8\pi v_d^2 \,,
\end{eqnarray}
where, we have used \Eqn{e:mA1} to substitute for $m_A$. One interesting thing to note from above inequality is that, when $v_t\ne 0$, the value of $m_H$ is bounded from above in the SM-like limit, $\sin\alpha = 0$. But what is more interesting is to find that,
for large $m_H$ and small $\sin\alpha$, the expression on the LHS of (\ref{e:peak}) vanishes for $\sin\alpha\approx 2v_t/v_d$ and so
the inequality is trivially satisfied. Thus the location of the peak on the horizontal axis of Fig.~\ref{f:samH} is a direct reflection
of a small value of $v_t$. Smaller the value for $v_t$, closer is the peak to $\sin\alpha=0$.

Strictly speaking, the peak in Fig.~\ref{f:samH} does not extend up to infinity along the vertical axis. This is because the unitarity
condition~(\ref{e:u3}) which in terms of physical masses reads
\begin{eqnarray}
\label{e:peak1}
\left(m_H^2\sin^2\alpha+m_h^2\cos^2\alpha \right) \le 16\pi v_d^2 \,,
\end{eqnarray}
always puts an upper bound on $m_H$ for nonzero $\sin\alpha$. But for $v_t\to 0$ the peak of Fig.~\ref{f:samH} at $\sin\alpha=0$
which corresponds to the decoupling limit as defined in \Eqn{case2} and in this case, the bound from (\ref{e:peak}) can be
alleviated and infinitely heavy nonstandard scalars can be allowed.

Fig.~\ref{f:mHmC} depicts that the splitting between $m_H$ and $m_+$ can be restricted from unitarity and stability. In Figs.~\ref{f:samH} and \ref{f:mHmC} we have also drawn the contours of $m_A$ and $m_{++}$ to emphasize that due to the
correlation between different parameters, experimental bound on any of the nonstandard masses can be translated into indirect
bounds on the other masses too. To illustrate, if we can rule out a doubly charged scalar below 500~GeV from direct searches at
the LHC, then, from Fig.~\ref{f:mHmC}, we also forbid a singly charged scalar below 350~GeV. Note that, these correlated bounds
do not crucially depend on the numerical value of $v_t$ as long as it is small. In passing we also remark that although the contours
of $m_{++}$ in Fig.~\ref{f:mHmC} have been drawn for $\sin\alpha=0$, they are not appreciably modified for $|\sin\alpha| < 0.2$.

Things become more interesting when we superimpose, in Fig.~\ref{f:mHmC}, the constraint arising from the electroweak $T$-parameter. For $v_t=1$~GeV or less, the major contribution to the $T$-parameter comes from the loops involving the new
nonstandard scalars. With $\sin\alpha\approx 0$ and $m_H\approx m_A$ (these two approximations can already be justified from
Fig.~\ref{f:samH}), the new physics contribution to the electroweak $T$-parameter is given by\cite{Lavoura:1993nq,Chun:2012jw}
\begin{eqnarray}
\label{e:t-par}
\Delta T = \frac{1}{4\pi \sin^2\theta_wm_W^2} \left[ F(m_+^2,m_A^2) + F(m_{++}^2,m_+^2) \right] \,,
\end{eqnarray}
where, $\theta_w$ and $m_W$ are the Weinberg angle and the $W$-boson mass respectively, and
\begin{eqnarray}
F(x,y) = \frac{x+y}{2}-\frac{xy}{x-y}\ln \left(\frac{x}{y} \right)\,.
\end{eqnarray}
In \Eqn{e:t-par} we further use the relation (\ref{e:mCC1}) to substitute for $m_{++}$. Taking the new physics contribution to the $T$-parameter as\cite{Agashe:2014kda}
\begin{eqnarray}
\Delta T < 0.2 ~~ {\rm at} ~~ 95\% {\rm C.L.} \,,
\end{eqnarray}
we draw the allowed region in Fig.~\ref{f:mHmC}. From there we see that the combined constraint implies that all the nonstandard scalars should be nearly degenerate.
 Using the correlation of \Eqn{e:mCC1}, and denoting the typical mass difference, $(m_+-m_{++})$ by $\delta$ ($\delta \ll m$), we can simplify the expression of the $T$-parameter as follows:
\begin{eqnarray}
\Delta T \approx \frac{\delta^2}{3\pi \sin^2\theta_wm_W^2}  \,.
\end{eqnarray} 
Using the experimental number we can then find $|\delta|\lesssim 50$~GeV.

\section{Impact on loop induced Higgs decays}\label{s:hgg}
Since the quarks couple only with the doublet and the physical scalar, $h$, in general, is a mixed state of doublet and triplet
fields, the tree level couplings of $h$ will be modified from those in the SM. We define a generic modification factor for the
fermions and vector bosons as follows:
\begin{eqnarray}
\kappa_X = \frac{g^{\rm model}_{hXX}}{g^{\rm SM}_{hXX}} \,.
\end{eqnarray}
Then, for $v_t\ll v_d$, one can easily calculate\cite{Arhrib:2011uy}
\begin{eqnarray}
\kappa_q \approx \kappa_V \approx \cos\alpha \,,
\end{eqnarray}
where, $q$ represents any fermion and $V=W, Z$. Denoting by $f_X$ the percentage of $h$ produced via the channel $X$, we
can express the modification of the Higgs production cross section as follows:
\begin{eqnarray}
{\cal R}_P =\frac{\sigma(pp\to h)^{\rm model}}{\sigma(pp\to h)^{\rm SM}} = \kappa_q^2 \left(f_{ggF}+f_{tth} \right) + \kappa_V^2 \left(f_{VBF}+f_{Vh} \right) \,,
\end{eqnarray}
where, $f_{ggF}\approx 87.5\%$, $f_{VBF} \approx 7\%$, $f_{Vh}\approx 5\%$ and $f_{tth}\approx 0.5\%$ at a CM energy
of 7 and 8~TeV\cite{Agashe:2014kda}. We know that a 125~GeV SM Higgs decays into $VV^*$ channel with nearly 24\%
branching ratio and the rest decays almost entirely into two-body fermionic channels and into two gluons. Thus the modification
of the total decay width can be expressed as
\begin{eqnarray}
{\cal R} =\frac{\Gamma^{\rm model}}{\Gamma^{\rm SM}} = \kappa_q^2 \cdot 76\% + \kappa_V^2 \cdot 24\% \,.
\end{eqnarray}

\begin{figure}[htbp!]
     	\centering
     	\begin{tabular}{c c}
     		\includegraphics[width=7cm,height=5cm]{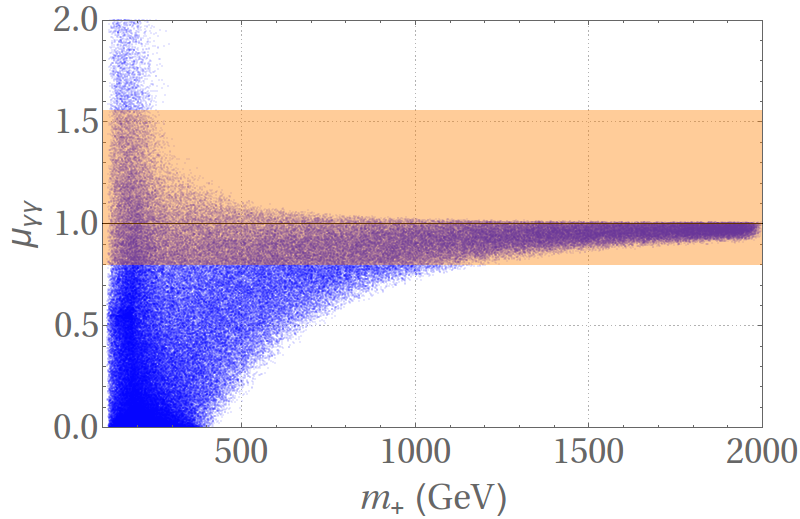} & 
     		\includegraphics[width=7cm,height=5cm]{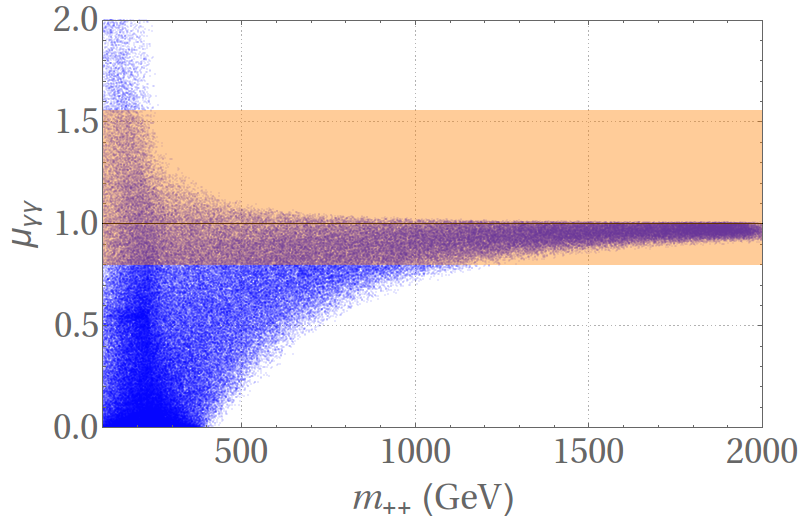} 
     	\end{tabular}
     	\caption{\em Dependence of the diphoton signal strength on $m_+$ and $m_{++}$ plotted for the points that survive
     	from the constraints of unitarity, stability and $T$-parameter~(blue points)  for $v_t=1$~GeV. The horizontal band represents the $2\sigma$ experimental limit from the combined fit of ATLAS and CMS results\cite{data}.}
     	\label{f:mCmu}
\end{figure}

Now we turn our attention to the modification of the partial decay widths of loop induced Higgs decays like $h\to \gamma\gamma$.
To display conveniently the contribution of the charged scalar loops to the decay amplitude, we define dimensionless parameters
$\kappa_+$ and $\kappa_{++}$ in the following way,
\begin{subequations}
\label{kappa}
\begin{eqnarray}
\kappa_+ &=& \frac{m_W}{gm_+^2} g_{hH^+H^-} \,, \\
\kappa_{++} &=& \frac{m_W}{gm_{++}^2} g_{h\delta^{++}\delta^{--}} \,,
\end{eqnarray}
\end{subequations}
where, the general expressions for $g_{hH^+H^-}$ and $g_{h\delta^{++}\delta^{--}}$ are given by
\begin{subequations}
\label{g}
\begin{eqnarray}
\label{g+}
g_{hH^+H^-} &=& -\frac{A_+\cos\alpha+ B_+\sin\alpha}{v_tv_d(v_d^2+2v_t^2)(v_d^2+4v_t^2)} \,, \\
{\rm with,} ~~ A_+ &=& 2v_t(v_d^2+4v_t^2)(m_+^2v_d^2+m_h^2v_t^2) \,, \\
B_+ &=& v_d\left\{(v_d^2+4v_t^2)(m_h^2v_d^2+4m_+^2v_t^2)-m_A^2(v_d^2+2v_t^2)^2 \right\} \,, \\
\label{g++}
{\rm and,}~~
g_{h\delta^{++}\delta^{--}} &=& -\frac{A_{++}\cos\alpha+ B_{++}\sin\alpha}{v_t(v_d^2+2v_t^2)(v_d^2+4v_t^2)} \,, \\
{\rm with,} ~~ A_{++} &=& 2v_tv_d\left\{2m_+^2(v_d^2+4v_t^2)-m_A^2(v_d^2+2v_t^2) \right\} \,, \\
B_{++} &=& \left[m_A^2v_d^2(v_d^2+2v_t^2)- (v_d^2+4v_t^2)\left\{4m_{+}^2-(2m_{++}^2 +m_h^2)(v_d^2+2v_t^2) \right\} \right] \,.
\end{eqnarray}
\end{subequations}
With these, the modification of the diphoton decay width can be written as
\begin{eqnarray}
\label{diphoton}
{\cal R}_{\gamma\gamma} =\frac{\Gamma(h\to \gamma\gamma)^{\rm model}}{\Gamma(h\to \gamma\gamma)^{\rm SM}} = \frac{\left|\kappa_V{\cal F}_1(\tau_W)+\frac{4}{3} \kappa_q{\cal F}_{1/2}(\tau_t) +\kappa_+{\cal F}_0(\tau_+) +4\kappa_{++}{\cal F}_0(\tau_{++}) \right|^2}{\left|{\cal F}_1(\tau_W)+\frac{4}{3} {\cal F}_{1/2}(\tau_t) \right|^2} \,,
\end{eqnarray}
where, using the notation, $\tau_x \equiv (2m_x/m_h)^2$, the ${\cal F}$ functions can be written as\cite{Gunion:1989we}
\begin{subequations}
\begin{eqnarray}
{\cal F}_1(\tau_x) &=& 2+3\tau_x +3\tau_x(2-\tau_x) {\cal G}(\tau_x) \,, \\
{\cal F}_{1/2}(\tau_x) &=& -2\tau_x\left\{1+(1-\tau_x) {\cal G}(\tau_x) \right\} \,, \\
{\cal F}_0(\tau_x) &=& -\tau_x\left\{1-\tau_x{\cal G}(\tau_x) \right\} \,, \label{e:f0} \\
{\rm with,}~~ {\cal G}(\tau) &=& \begin{cases}
\left[\sin^{-1}\left(\sqrt{\frac{1}{\tau}} \right) \right]^2 \,, & \text{for } \tau\ge 1 \,, \\
-\frac{1}{4}\left[\ln\left(\frac{1+\sqrt{1-\tau}}{1-\sqrt{1-\tau}}\right)-i\pi \right]^2 \,, & \text{for } \tau < 1 \,.
\end{cases}
\end{eqnarray}
\end{subequations}
Now we can write the modified Higgs signal strength as follows:
\begin{eqnarray}
\label{e:mugg}
\mu_{\gamma\gamma} = \frac{{\cal R}_P}{{\cal R}} \times {\cal R}_{\gamma\gamma} \,.
\end{eqnarray}
In Fig.~\ref{f:mCmu} we have shown how $\mu_{\gamma\gamma}$ behaves with the charged scalar masses. Here, one should note
that the signal strength is suppressed compared to the SM expectations for heavy charged scalars. Thus observation of
an excess in the diphoton channel, in future runs of the LHC, will disfavor the possibility of heavy charged scalars in this model. In passing,
we comment that although we have assumed $v_t=1$~GeV in Fig.~\ref{f:mCmu}, the above conclusions do not crucially
depend on our numerical choice of $v_t$ as long as it remains small.

\begin{figure}[htbp!]
\begin{minipage}{0.46\textwidth}
\centerline{\includegraphics[width=8cm,height=6cm]{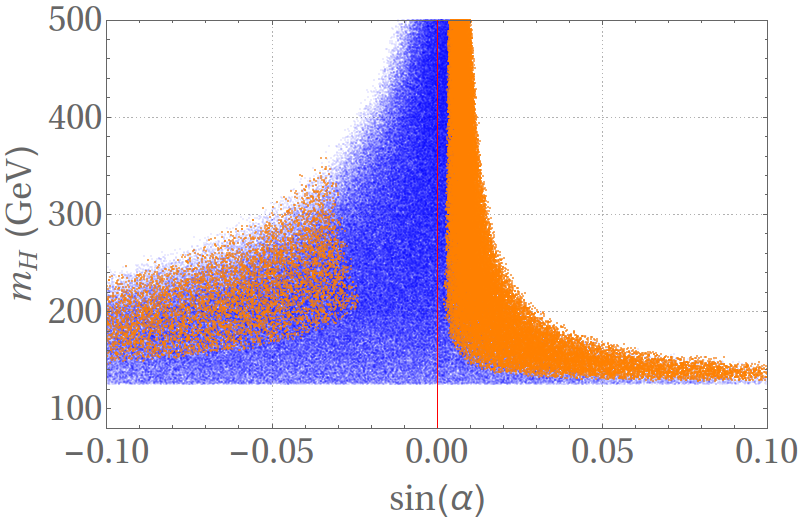}}
\caption{\em The blue points (in the background) are allowed from unitarity, stability and $T$-parameter for $v_t=1$~GeV. The yellow points are those which survive when 
$2\sigma$ constraint from $\mu_{\gamma\gamma}$ are added on top of it. Clearly, a narrow band around $\sin\alpha\approx 0$ can be ruled out from $\mu_{\gamma\gamma}$.}
\label{f:overlap}
\end{minipage}
\hfill
\begin{minipage}{0.46\textwidth}
\centerline{\includegraphics[width=8cm,height=6cm]{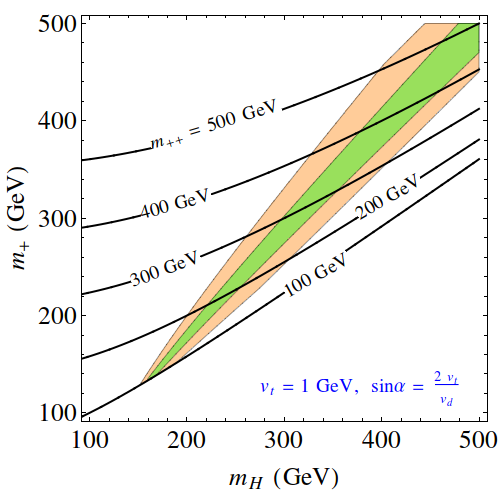}}
\caption{\em The outer orange region is allowed from $T$-parameter and the observed diphoton signal strength at 95\% CL. The inner green region represents how the constraint will tighten if $\mu_{\gamma\gamma}$ is measured to be $1\pm 5\%$. Contours of  $m_{++}$ are also shown.}
  \label{f:diphoton}
\end{minipage}
\end{figure} 

Next, we can also define, similar to \Eqn{e:mugg}, the modified signal strength for the $h\to Z\gamma$ decay channel,
$\mu_{Z\gamma}$. However, for brevity, we do not display the explicit expressions here. Interested readers can find the
relevant formulas in some earlier papers\cite{Arbabifar:2012bd, Dev:2013ff, Chen:2013dh,Haba:2016zbu}. As has been noted in these
references, since the sign of the $H^+H^-Z\gamma$ vertex is opposite to that of the $\delta^{++}\delta^{--}Z\gamma$
vertex, the singly charged scalar loop interfere destructively with the doubly charged one in the $h\to Z\gamma$ amplitude.
Consequently, in some region of the parameter space depending on which loop dominates, we can have an {\em anti-correlation}  between $\mu_{\gamma\gamma}$ and $\mu_{Z\gamma}$, {\em i.e.}, one is enhanced compared to the SM while the other is suppressed.

But the new thing that we want to add to this context is the understanding of the behavior of $\mu_{\gamma\gamma}$
and $\mu_{Z\gamma}$ in the SM-like scenario. One should note that, as the charged scalars become heavy, the function
${\cal F}_0(\tau)$ in \Eqn{e:f0} saturates to $1/3$. Hence the decoupling of the charged scalars from the loop induced Higgs
decays depends on how $\kappa_+$~($\kappa_{++}$) behaves with increasing $m_+$~($m_{++}$). We can easily check that,
in the SM-like case defined by \Eqn{case1}, the trilinear couplings in \Eqs{g+}{g++} take the following forms:
\begin{subequations}
\label{glim}
\begin{eqnarray}
\label{g1+}
g_{hH^+H^-} &=& -\frac{2\left(m_+^2v_d^2+m_h^2v_t^2\right) }{v_d(v_d^2+2v_t^2)} \approx -\frac{2m_{+}^2}{v_d} \,, \\
\label{g1++}
{\rm and,}~~
g_{h\delta^{++}\delta^{--}} &=& -2v_d\left[\frac{2m_+^2}{v_d^2+2v_t^2}-\frac{m_A^2}{v_d^2+4v_t^2} \right] \approx -\frac{2m_{++}^2}{v_d}  \,,
\end{eqnarray}
\end{subequations}
where, in the final step we have used \Eqs{e:mA1}{e:mCC1}. Thus it follows from \Eqn{kappa} that $\kappa_+, \kappa_{++} \to -1$
when $m_+, m_{++} \gg m_h$.  Consequently the
charged scalars contribute substantially to the diphoton or $Z$-photon decay amplitudes even when their masses lie in the TeV
regime. In fact, one can check from \Eqn{e:mugg} that $\mu_{\gamma\gamma} \approx {\cal R}_{\gamma\gamma}\lesssim 0.5$
when $\sin\alpha=0$ for finite $v_t$. Therefore the SM-like limit defined by \Eqn{case1} is already forbidden from the current
combined fit value of the diphoton signal strength, $\mu_{\gamma\gamma}=1.16^{+0.20}_{-0.18}$\cite{data}. This feature has been clearly depicted in Fig.~\ref{f:overlap} where we can see that the region around $\sin\alpha=0$ is ruled out from the Higgs to diphoton data. The reason for this large quantum effect for heavy masses
 can be understood once we realize that both the VEVs in this case are nonzero. This means the charged
scalars obtain their masses entirely from spontaneous symmetry breaking (SSB), which leads to large quantum effects as also has been
noted in the multi doublet context\cite{Bhattacharyya:2014oka}.

But in the SM-like limit defined by \Eqn{case2} one can obtain
\begin{subequations}
\label{glim2}
\begin{eqnarray}
\label{g2+}
g_{hH^+H^-} &=& -\frac{2\left(m_+^2-m_A^2\right)+2m_h^2 }{v} \,, \\
\label{g2++}
{\rm and,}~~
g_{h\delta^{++}\delta^{--}} &=& -\frac{2\left(m_{++}^2-m_A^2\right)+2m_h^2}{v}   \,.
\end{eqnarray}
\end{subequations}
Here we can easily see that the charged scalars can be decoupled in the limit,
\begin{eqnarray}
\label{dec}
m_+\approx m_{++} \approx m_A \gg m_h \,.
\end{eqnarray}
Let us investigate the above limit in some greater detail. Since $v_t\to 0$ in this case, the triplet remains inert and
$M^2$ in \Eqn{e:potential} serves as a free mass squared parameter not connected to SSB. Using \Eqn{e:M2} to recover $M^2$
from $v_t$ and then plugging it in \Eqn{e:mu} we can see that for this type of scenario, $m_A^2\approx M^2$. Thus the
condition~(\ref{dec}) essentially implies that the charged scalars obtain all their masses from a non-SSB origin, which, not
surprisingly, lead to decoupling in the same way as in the inert doublet models\cite{Bhattacharyya:2014oka}.

From the combined region in Fig.~\ref{f:overlap}, we see that the space for $m_H$~(or, $m_A$) is severely constrained ($m_H<340$~GeV) when the value of $\sin\alpha$ departs from $2v_t/v_d$. As we will discuss below, even in the region $\sin\alpha\approx 2v_t/v_d$
we can obtain lower bounds on masses that are competitive with the direct experimental bounds.
%

%
Much attention has been received by the doubly charged scalars because of the possibility of detecting them in the same sign
dilepton channel. In this article, we have shown that unitarity and $T$-parameter impart a degeneracy among the nonstandard
scalars for $v_t\lesssim 1$~GeV. Because of this, the doubly charged scalar can now decay into mainly three channels\cite{Melfo:2011nx} -- (i)~same
sign dileptons, (ii)~a pair of same sign $W$-bosons and (iii)~a $W$-boson and a singly charged scalar. 
As we will explain shortly, continuation of the agreement of the diphoton signal strength with the corresponding SM
expectation will result in tightening the degree of degeneracy between the nonstandard scalar masses. In that case, for the usual type~II scenario, the 
decay mode $\delta^{\pm\pm}\to W^\pm H^\pm$ is likely to be suppressed and the doubly charged scalar will decay almost
exclusively into same sign dileptons and/or same sign $W$-bosons.
When $\delta^{++}$ decays completely into two same sign dileptons,
it is easy to look for it in the experiments and the bound on its mass is quite strong ($m_{++}\gtrsim 400$~GeV)\cite{CMS:2016cpz}.
In the type~II seesaw model, the dominant decay mode of $\delta^{++}$ depends on the VEV of the triplet\cite{Han:2007bk,Garayoa:2007fw,Kadastik:2007yd,Akeroyd:2007zv}.
For $v_t\lesssim\order(10^{-4})$~GeV, the doubly charged scalar decays entirely into a pair of leptons and the bound from direct
searches applies.  But in the region
$\order(10^{-4})\lesssim v_t\lesssim\order(1)$~GeV, the $\delta^{++}$ decays mainly into a pair of $W$-bosons and
therefore, it is very difficult to search for. In this case, a very weak bound has been placed on the mass of the doubly charged scalar using the LHC data\cite{Kanemura:2013vxa,Kanemura:2014ipa,kang:2014jia}. In this context, we note that too low values for the
masses of the charged scalar may give substantial contribution to the diphoton decay amplitude and therefore, it might be possible
to set a lower bound on the masses from the observed value of the diphoton signal strength. Since all the nonstandard scalar
masses are correlated, this bound can be translated into bounds on other scalar masses also. Taking into account the experimental bound from LEP-2, $m_{++}>100$~GeV\cite{Abdallah:2002qj}, along with the
mass relations of \Eqs{e:mA1}{e:mCC1} we plot the $2\sigma$ allowed region from $T$-parameter and the observed diphoton
signal strength in Fig.~\ref{f:diphoton}. Although we have displayed our results for $v_t=1$~GeV, the plot remains essentially the 
same for any value of $v_t\lesssim 1$~GeV. From Fig.~\ref{f:diphoton}, we read the following lower bounds on the nonstandard
scalar masses: $m_{+}\gtrsim 130$~GeV, $m_{H}\approx m_A \gtrsim 150$~GeV. In particular,
for $v_t\lesssim\order(10^{-4})$~GeV when $m_{++}>400$~GeV applies, we can give the following lower bounds on the other
nonstandard scalar masses: $m_{+}\gtrsim 365$~GeV, $m_{H}\approx m_A \gtrsim 330$~GeV. From \Eqs{g2+}{g2++} we see that
a more precise measurement of $\mu_{\gamma\gamma}$ has the potential to constrain the mass splittings more strongly than the 
$T$-parameter. To illustrate this point, in Fig.~\ref{f:diphoton}, we have also showed the futuristic projection of the $2\sigma$ allowed parameter
region (in green) assuming that the diphoton signal strength will be measured to be consistent with the SM within 5\% accuracy level. In that
case, the bounds for $v_t\lesssim\order(10^{-4})$~GeV can be improved to $m_{+}\gtrsim 390$~GeV, $m_{H}\approx m_A \gtrsim 375$~GeV.

\section{Conclusions}
\label{s:conclusions}
In this paper we have revisited the constraints on the scalar sector of the type~II seesaw model. We have worked under the assumption
that the lightest CP-even scalar~($h$) has been observed the LHC. Although we have exemplified our results for $v_t=1$~GeV, our conclusions
are mostly generic and valid for any $v_t$ of $\order(1~{\rm GeV})$ or less. To begin with, we have re-derived the correlations between
the nonstandard masses using unitarity. We have expressed all our results in terms of physical masses and mixing angles. Consequently,
we have noticed that $\sin\alpha$ becomes restricted within a narrow range around $2v_t/v$ just from unitarity, whenever the neutral
nonstandard scalar is heavier than 300~GeV. Moreover, for nonzero $v_t$, we have argued how a thin strip around $\sin\alpha=0$ can be ruled out from
$h\to\gamma\gamma$, which shrinks further the allowed band for $\sin\alpha$ (around $2v_t/v$). Using the mass relations in conjunction with
the $T$-parameter, the separation between the masses, $|m_+-m_{++}|\approx |m_+-m_A|$, can be restricted to be less than 50~GeV. We have also found that, for
$m_{++}>700$~GeV, an enhancement in $\mu_{\gamma\gamma}$ is hardly possible. Finally, using the experimental range of
$\mu_{\gamma\gamma}$ and $T$-parameter we plotted the $2\sigma$ allowed region in the $m_H$-$m_+$ plane for $\sin\alpha=2v_t/v$.
Using $m_{++}>100$~GeV from LEP-2, we obtain the following limits on the other masses:
\begin{equation*}
m_+ > 130~{\rm GeV} \,, ~~~ m_{H,A} > 150~{\rm GeV}\,.
\end{equation*}
Moreover, when $v_t<10^{-4}$~GeV, the direct search bound from LHC, $m_{++}>400$~GeV applies and then we can improve the
bounds on other masses as follows:
\begin{equation*}
m_+ > 365~{\rm GeV} \,, ~~~ m_{H,A} > 330~{\rm GeV}\,.
\end{equation*}
Remembering that $T$-parameter restricts the mass splitting, we may conclude that the spectrum can be determined by essentially one
mass parameter for $v_t<10^{-4}$~GeV. We have also hinted how, in future, $\mu_{\gamma\gamma}$ can play a very important role
in restricting the splittings between nonstandard masses.

\section*{Acknowledgements} 
We thank Ipsita Saha for useful discussions.
This work has been partially supported by the Spanish MINECO under grants  FPA2011-23897, FPA2014-54459-P, by the ``Centro de Excelencia Severo Ochoa'' Programme under grant SEV-2014-0398 and by the ``Generalitat Valenciana'' grant GVPROMETEOII2014-087.



\providecommand{\href}[2]{#2}\begingroup\raggedright\endgroup

\end{document}